\documentclass[10pt,conference]{IEEEtran}
\IEEEoverridecommandlockouts

\usepackage{graphicx} 
\usepackage{url}
\usepackage{listings}
\usepackage{xcolor}
\usepackage{amsmath}
\usepackage{amssymb}
\usepackage{flushend}
\usepackage{balance}

\graphicspath{{figs/}}

\usepackage{ifthen}
\newboolean{showcomments}
\setboolean{showcomments}{false}          
\ifthenelse{\boolean{showcomments}}
  {\newcommand{\nb}[2]{
    \fcolorbox{gray}{yellow}{\bfseries\scriptsize#1}
    {\scriptsize$\blacktriangleright${#2}$\blacktriangleleft$}
   } 
  }
  {\newcommand{\nb}[2]{}}

\makeatletter
\newcommand{\linebreakand}{%
  \end{@IEEEauthorhalign}
  \hfill\mbox{}\par
  \mbox{}\hfill\begin{@IEEEauthorhalign}
}
\makeatother

\title{Towards Automated Governance: A DSL for Human-Agent Collaboration in Software Projects%
\thanks{This work has been funded by the European Union under the Grant Agreement No 101189664 (MOSAICO project) and TED2021-130331B-I00 funded by MCIN/AEI/10.13039/501100011033 and European Union NextGenerationEU/PRTR. Views and opinions expressed are those of the author(s) only and do not necessarily reflect those of the European Union or the European Health and Digital Executive Agency (HADEA). Neither the European Union nor the granting authority can be held responsible for them.
Jordi Cabot is supported by the Luxembourg National Research Fund (FNR) PEARL program, grant agreement 16544475.}}

\date{June 2025}

\author{
    \IEEEauthorblockN{Adem Ait}
    \IEEEauthorblockA{\textit{University of Luxembourg}\\
    Esch-sur-Alzette, Luxembourg\\
    adem.ait@uni.lu}
    \and
    \IEEEauthorblockN{Gwendal Jouneaux}
    \IEEEauthorblockA{\textit{Luxembourg Institute of Science and Technology}\\
    Esch-sur-Alzette, Luxembourg \\
    gwendal.jouneaux@list.lu}
    \linebreakand    
    \IEEEauthorblockN{Javier Luis C\'anovas Izquierdo}
    \IEEEauthorblockA{\textit{Universitat Oberta de Catalunya} \\
    Barcelona, Spain \\
    jcanovasi@uoc.edu}
    \and 
    \IEEEauthorblockN{Jordi Cabot}
    \IEEEauthorblockA{\textit{University of Luxembourg} \\
    \textit{Luxembourg Institute of Science and Technology}\\
    Esch-sur-Alzette, Luxembourg \\
    jordi.cabot@list.lu}
}

\begin{document} 

\maketitle

\begin{abstract}
    The stakeholders involved in software development are becoming increasingly diverse, with both human contributors from varied backgrounds and AI-powered agents collaborating together in the process.
    This situation presents unique governance challenges, particularly in Open-Source Software (OSS) projects, where explicit policies are often lacking or unclear.
    This paper presents the vision and foundational concepts for a novel Domain-Specific Language (DSL) designed to define and enforce rich governance policies in systems involving diverse stakeholders, including agents. 
    This DSL offers a pathway towards more robust, adaptable, and ultimately automated governance, paving the way for more effective collaboration in software projects, especially OSS ones.
\end{abstract} 

\section{Introduction}
\label{sec:introduction}
Governance is a key aspect of software development, especially in Open-Source Software (OSS), where the collaborative nature of the process requires clear guidelines and policies to ensure effective decision-making and accountability.
Governance policies help to establish roles, responsibilities, and processes for managing contributions, resolving conflicts, and maintaining the integrity of software projects~\cite{DBLP:journals/cacm/IzquierdoC23}.
Such policies make explicit the decision-making process, ensuring that all the involved stakeholders know how decisions are made and who is responsible for them.
However, most (OSS) projects do not explicitly define governance policies, and those that do are often poorly described and lack clarity~\cite{DBLP:journals/cacm/IzquierdoC23}.
Furthermore, when described, they are usually scattered across different project resources, making it difficult for contributors to understand the rules and procedures that govern the project~\cite{DBLP:conf/icse/IzquierdoC15}.

Recently, the landscape of software development is undergoing a profound transformation.
The rapid proliferation of AI-powered agents participating in development tasks, coupled with a growing recognition of the critical role of diverse human backgrounds, presents unprecedented challenges to these established governance paradigms~\cite{DBLP:journals/pacmhci/WesselSSWPCG18, DBLP:conf/xpu/RasheedWSKA0SA24, bjorn2023diversity, DBLP:journals/ese/Rodriguez-Perez21}.
AI agents, powered by advancements in Large Language Models (LLMs), are moving beyond simple automation to become active participants capable of complex communication, collaboration, and even decision-making~\cite{DBLP:conf/ijcai/GuoCWCPCW024}.
On the other hand, human diversity is also beneficial for software development, as it can lead to more innovative solutions and better decision-making~\cite{yang2022gender}, while also benefiting end-users~\cite{gunatilake2024impact}.
While this human-agent collaboration and broader human diversity promise significant benefits in innovation and productivity, they also expose a critical gap: the lack of governance frameworks designed to explicitly and holistically manage such multifaceted participation.\looseness-1

This paper first replicates a study on governance policies in OSS~\cite{DBLP:conf/icse/IzquierdoC15} to demonstrate the persistent need for explicit governance, particularly in today's agentic development context.
Then, we address this gap by proposing the conceptualization of a Domain-Specific Language (DSL) designed to define and enforce governance policies in systems involving a diverse set of stakeholders, covering both human and agent collaborators. 
The proposed DSL provides a structured approach for expressing governance rules, roles, responsibilities, and decision-making processes.
It encompasses various dimensions of governance, including decision-making procedures, positive and negative discrimination, and other methods to ensure a fair representation.
We envision our DSL to be field-agnostic, extensible, and adaptable to different contexts, allowing for the inclusion of various stakeholders and their specific needs.
We also provide tool support to automate and enforce the governance policies defined with our DSL.

The rest of the paper is structured as follows. 
Section \ref{sec:motivation} provides the motivation.
Section~\ref{sec:dsl} and \ref{sec:proof-of-concept}  presents the design and implementation of the DSL, and the proof of concept, respectively.
Section~\ref{sec:discussion} discusses the roadmap.
Finally, Section~\ref{sec:conclusion} concludes the paper and outlines the future work.


\section{State of the Art}
\label{sec:motivation}

\subsection{Diversity and Governance in Software Engineering}

Software engineering teams are becoming increasingly diverse across multiple dimensions.
Recent research has highlighted the importance of diversity within software development processes~\cite{bjorn2023diversity,DBLP:journals/ese/Rodriguez-Perez21}.
Diversity is being studied in terms of gender, race or ethnicity, but also in terms of cognitive diversity~\cite{DBLP:journals/corr/abs-2503-05470,DBLP:conf/icse/DuttaCST23}.
Studies have shown that diverse teams tend to produce more innovative solutions~\cite{DBLP:conf/icse/CatolinoPTSF19,doi:10.1073/pnas.2200841119}, and impact positively in software productivity~\cite{DBLP:conf/chi/VasilescuPRBSDF15}.

Adding to this human diversity, software projects now include a growing number of non-human participants in the form of bots and AI-powered agents~\cite{DBLP:journals/pacmhci/WesselSSWPCG18,DBLP:conf/xpu/RasheedWSKA0SA24}.
These automated participants have traditionally performed routine tasks like continuous integration, code quality checks, and dependency management~\cite{DBLP:conf/msr/KinsmanWGT21}.
While these tools have traditionally handled routine tasks and basic governance functions~\cite{DBLP:journals/ese/WesselVGT23,DBLP:conf/icsm/DecanMMG22}, 
the rise of LLMs is transforming these agents from mere automation tools into sophisticated collaborators capable of reasoning, communication, and participation in complex decision-making~\cite{xi2025rise, DBLP:conf/ijcai/GuoCWCPCW024}. 
This profound shift introduces an unprecedented dimension of diversity, presenting novel and urgent conceptual challenges for governance that existing frameworks are missing.

\subsection{Governance in practice: the case of OSS development}
\label{sec:motivation:governance-oss}

A previous study illustrated the need for explicit governance policies in OSS~\cite{DBLP:conf/icse/IzquierdoC15} but lacked the mechanisms to address the challenges introduced by participant diversity, particularly the inclusion of AI agents in decision-making processes.
Our current proposal aims to fill this gap by incorporating diversity-aware governance mechanisms that acknowledge the different characteristics of participants and ensure equitable representation in decision-making processes.

While the landscape of participants is evolving, the explicitness and nature of governance policies in practice, particularly in OSS, remain a concern. 
To understand this current state, we replicated the study conducted by C\'anovas Izquierdo and Cabot
~\cite{DBLP:journals/cacm/IzquierdoC23}, which analyzed the top 25 starred software projects and found that most OSS projects do not have explicit governance policies, and those that do are often poorly defined and lack clarity.
The governance evidence is analyzed from the explicit indications of four dimensions: 
(1) whether there is some workflow to contribute, different from the typical pull-based development process; 
(2) who makes the decisions to accept code, and how; 
(3) how long it takes to review or accept a contribution; 
and (4) how to become a contributor.\looseness-1


We found that 68\% of repositories reported at least one of the four dimensions, 24\% reported none, and 8\% reported all four dimensions. 
The situation is showing slight improvements from the previous study.
In comparison, we have seen an increase in the number of projects that partially discussed governance, from 32\% to 68\%, mainly because of the report of the contribution process.
However, the span to review or accept a contribution is only defined in 16\% of the projects, how to become a contributor is only defined in 20\% of the projects, and who makes the decisions, and how, to accept code is only defined in 24\% of the projects.
Only 2 projects (8\%) provided a full description of the policies governing them. 
Note that no project included a governance.md file but reported them in the contributing.md file or on their documentation website.
None of the analyzed projects adopted a DSL to specify their governance policies, which could help in the definition, enforcement, and automation of the governance policies~\cite{DBLP:journals/cacm/IzquierdoC23}.

\section{Defining Governance Policies}
\label{sec:dsl}

A DSL can formalize governance policies, roles, and decision-making procedures, enhancing transparency and reliability.
To the best of our knowledge, there is no DSL for defining governance policies in systems that address the diversity of participants.

Several studies 
analyze the governance of software projects~\cite{DBLP:conf/icse/ChulaniWY08,DBLP:conf/sigsoft/Herbsleb16}, particularly 
OSS projects~\cite{DBLP:journals/ism/PeltJBO21,DBLP:conf/icse/IzquierdoC15,DBLP:conf/ease/KeertipatiLS16}.
However, to the best of our knowledge, only C\'anovas Izquierdo and Cabot~\cite{DBLP:conf/icse/IzquierdoC15} proposed a DSL, where they covered the main dimensions  
to define and enforce governance rules in OSS projects.
Note, their proposal did not provide support to assess the diversity of the participants involved in the decision-making process.
Our proposal completely revamps that DSL and adapts it to the current reality of software development involving participants with different profiles, the possibility of having uncertain AI agents, and much more complex decision procedures with different participant weighting systems.

To collect the requirements for our DSL, we have carefully considered existing AI governance frameworks and principles~\cite{DBLP:journals/computer/ChestermanGHS24}. 
For instance, we incorporate key AI governance indicators identified in the literature, such as autonomy level~\cite{DBLP:journals/nature/RahwanCOBBBCCCJ19} and explainability~\cite{DBLP:journals/inffus/ArrietaRSBTBGGM20}%
, thus integrating them directly into our language constructs for agent representation.

Based on these frameworks and our own experience in open-source development and the analysis of open-source communities, we propose, in what follows, our governance DSL.
A DSL is composed of three main elements~\cite{DBLP:books/sp/WasowskiB23}: 
(1) abstract syntax, which defines the concepts and relationships of the domain where the language is applied; 
(2) concrete syntax, which defines the notation of the language (e.g., textual, diagram-based, etc.); and 
(3) semantics, which defines the meaning of the language constructs.
In the following, we discuss each component. 

\subsection{Abstract Syntax}
\label{sec:our-dsl:abstract-syntax}

\begin{figure*}[t]
    \centering
    \includegraphics[width=\textwidth]{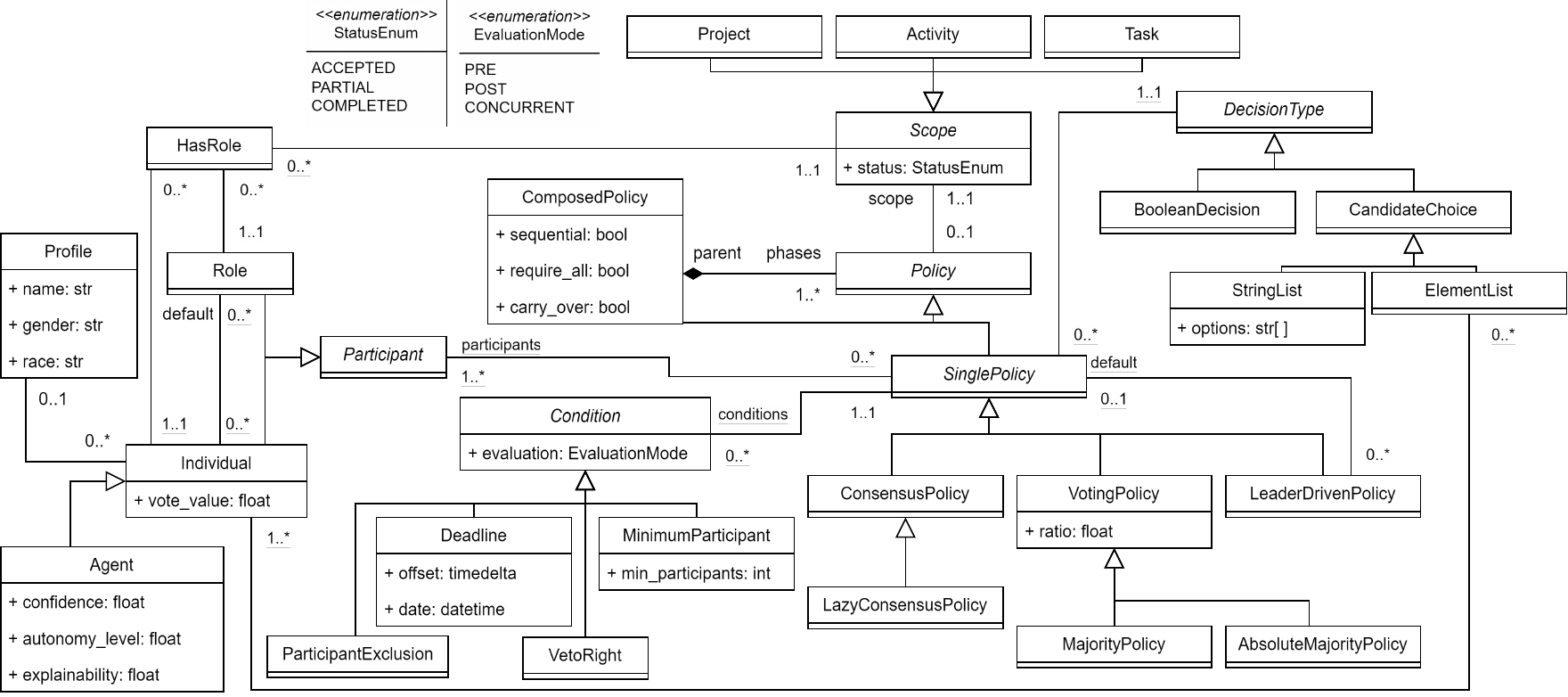}
    \caption{Abstract syntax metamodel of our DSL.}
    \label{fig:metamodel}
\end{figure*}

The abstract syntax of our DSL is defined via the metamodel shown in Figure~\ref{fig:metamodel}. 
Metamodels restrict the structure of valid models (i.e., DSL instances, in our case, concrete policies) and define the relationships between the different elements of the language (i.e., the different concepts that can be used to define a policy and how these concepts can be linked to each other)~\cite{DBLP:series/synthesis/2017Brambilla}.
Governance models conforming to this metamodel represent specific sets of governance policies.

Our proposal is structured around key conceptual areas:

\smallskip
\noindent\textbf{Participants: Embracing Diversity and Agent Characteristics.}
A vital aspect of our DSL is the representation of \texttt{Participant}. 
The metamodel distinguishes between human participants (potentially detailed with \texttt{Profile} information like gender or role-specific attributes to support fairness or weighted influence) and \texttt{Agent}s. 
The influence of each participant in the decision-making process can be adjusted through the \texttt{vote\_value} attribute, which can be set to favor certain individuals or groups.
For \texttt{Agent}s, we describe further attributes critical for governance, such as \texttt{autonomy\_level}, \texttt{explainability} and \texttt{confidence}. 
This allows expressing complex policies, such as giving a high-autonomy agent more freedom, while decisions from agents with low explainability might be used only as support.
Participants can also be grouped by \texttt{Role}, while also being flexible to allow certain individuals to act as part of a role in specific policies (see \texttt{hasRole}).\looseness-1

\smallskip
\noindent\textbf{Policies: Defining Flexible and Composable Decision-Making.}
The DSL provides a rich set of \texttt{Policy} constructs to capture diverse decision-making processes essential for adaptive governance.
It supports various policy resolution strategies, including \texttt{VotingPolicy} (e.g., majority and qualified majority based on a \texttt{ratio}), \texttt{ConsensusPolicy} (including \texttt{LazyConsensus}~\cite{LAZYCONSENSUSWEBSITE} as a practical mechanism for efficient agreement), and \texttt{LeaderDrivenPolicy} (with optional fallback mechanisms). 
Policies specify a \texttt{DecisionType}, such as \texttt{BooleanDecision} (for accept/reject scenarios like pull request approvals) or \texttt{CandidateChoice} (for selecting one out of multiple options, like electing a leader).
Simple policies can be combined into \texttt{ComposedPolicy} structures (e.g., requiring sequential evaluation or all sub-policies to pass). 
This enables the modular construction of complex governance scenarios from simpler, reusable components, thus enhancing clarity and maintainability.

\smallskip
\noindent\textbf{Scope and Conditions: Contextualizing Policy Application.}
Governance policies are usually applied in a particular project or granular component of it. Our DSL allows their application to be precisely contextualized.
Policies are applied within a defined \texttt{Scope}, such as a \texttt{Project}, an \texttt{Activity} (e.g., development, testing), or a specific \texttt{Task} (e.g., a patch). 
This hierarchical scoping enables granular control, allowing different governance rules for different project areas or activities.
Furthermore, policies can be described with \texttt{Condition}s (e.g., setting a deadline or a minimum required participants). 
These conditions can also act as prerequisites (\texttt{pre}) or checks following execution (\texttt{post}).

\subsection{Concrete Syntax}
\label{sec:our-dsl:concrete-syntax}

As concrete syntax, we use a textual notation following a typical block-based structure (see Listing~\ref{lst:example}).
The syntax is therefore formally defined by a grammar. 
Each instance of the metaclass is textually represented by its keyword and a block that contains the properties of the instance.
Containment references are represented as nested blocks while non-containment references use an identifier to refer to the target element.\looseness-1

\begin{lstlisting}[caption=Simple governance policy evaluated following a majority strategy as a \texttt{BooleanDecision} requiring 40\% positive votes with a deadline of ten days., 
                    label=lst:example, 
                    basicstyle=\ttfamily\fontsize{8pt}{8.5pt}\selectfont,
                    breaklines=true,
                    float=t,
                    captionpos=b]
Scopes: 
    Project myProject {
        activities : myActivity {
            tasks : myTask
        }
    }
Participants:
    Profiles : 
        profile1 {
            gender : male
            race : hispanic
        }
    Roles : Maintainer
    Individuals : 
        Joe {
            vote value : 0.7
            profile : profile1
            role: Maintainer
        }, 
        (Agent) Mike {
            confidence : 0.8
            role: Maintainer
        },
        Paul {
            role: Maintainer
        }
MajorityPolicy TestPolicy {
    Scope: myTask
    DecisionType as BooleanDecision
    Participant list : Maintainer
    Conditions:
        Deadline : 10 days
        ParticipantExclusion : Paul
    Parameters:
        ratio : 0.4
}
\end{lstlisting}
\subsection{Operational Semantics: Decision engine}
The operational semantics take the form of a decision engine, enforcing the policies on ongoing collaborations.
The engine receives platform data to update the decision-making process state and take the necessary actions.
For the first implementation of our approach, we rely on \textsc{GitHub} as the collaboration platform.

The decision engine operates through a continuous cycle of state management and policy enforcement. 
When a collaboration event occurs (e.g., a pull request is created), the engine captures the event data and creates an internal representation. 
This state management enables the engine to track multiple concurrent decision-making processes.\looseness-1

For new collaborations, the engine identifies applicable policies based on scope.
Once found, the engine creates an internal deadline check based on the policy to resolve the 
process.
When voting occurs (e.g., a pull request review), 
eligible users' votes are registered in the collaboration's ballot box.
Each vote is captured with its rationale, timestamp, and association to the voter. 
At deadline, the engine analyzes the votes 
and enacts the decision (e.g., merge the pull request).
Composed policies resolve phases
sequentially or in parallel, 
combining decisions via conjunction or disjunction.
In addition, 
votes can be carried over between phases.

\section{Proof of Concept}
\label{sec:proof-of-concept}
We implemented the metamodel as a set of Python classes, while the concrete syntax was implemented through a grammar definition using ANTLR~\cite{ANTLRWEBSITE}, 
a parser generator tool.
We developed a transformation that converts the Abstract Syntax Tree (AST) generated by ANTLR into instances of our metamodel, which serve as input for the decision engine.
The complete implementation, including the grammar and decision engine, are available in the tool repository.\footnote{\url{https://doi.org/10.5281/zenodo.15856633}}
We also translated several governance policies identified in our background analysis (see Section~\ref{sec:motivation:governance-oss}) into our DSL syntax to demonstrate its expressiveness and practical applicability.




\section{Roadmap}
\label{sec:discussion}
The proposed DSL is just a first step towards a more diverse,  
transparent and agentic collaboration in software development. 
Open challenges and future directions are discussed below.

\smallskip
\noindent\textbf{Usability improvement via new syntaxes.}
To facilitate the adoption of our DSL by less technical users, we envision offering alternative UIs, like chatbots, to enhance usability.
This will be especially useful when introducing our DSL to manage governance in other, non-software, domains.

\smallskip
\noindent\textbf{Longitudinal study.}
To better understand if explicit governance rules help to increase 
OSS contributions,
we plan a longitudinal study where we could compare the evolution of contributions in projects before and after they committed to an explicit governance model. 
This same study could be useful to see whether projects are sticking to such governance model or they just ignore it even after committing to it. 


\smallskip
\noindent\textbf{Improved impact of Multi-agent systems in software development.}
We plan to test whether our governance layer enables more complex multi-agent systems to collaborate, in a controlled way, to solve more ambitious software engineering tasks. 
By leveraging SWE-bench~\cite{DBLP:conf/iclr/JimenezYWYPPN24}, we can systematically evaluate how governance policies affect agent collaboration, task completion rates, and adherence to defined constraints.

\smallskip
\noindent\textbf{Participant identification and bias mitigation.}
A problem some projects face is identifying the right people to play certain roles. 
Given a governance model, it would be possible to analyze past project data to propose candidates to fill the roles required in the governance model. 
A real-time monitoring component should also be helpful to proactively change the candidates to continuously improve the diversity of the collaborations. Indeed, a key element of this (semi)automatic participant analysis could be devoted to make sure the participants list is not biased against certain communities and that it respects the profile requirements. 


\smallskip
\noindent\textbf{Governance beyond software.}
We believe other domains could also benefit from an explicit and precise definition of their governance policies based on our DSL. 
For instance, making explicit the governance systems of organizations (such as NGOs) would enable systematic comparison of their decision-making structures and their assessment of inclusivity. 

\section{Conclusion}
\label{sec:conclusion}
We have presented a vision of a DSL for the explicit definition and automation of governance policies in software projects, with a particular focus on supporting diversity among participants, including both human contributors and AI-powered agents. 
Our DSL extends previous work by enabling the specification of participant profiles and agent attributes, thus facilitating more equitable and transparent decision-making processes. 
This is just a first step in this direction. As future work, we plan to work on the aforementioned roadmap.

\bibliographystyle{ieeetr}
\bibliography{ase-main}

\end{document}